\documentclass[a4paper,12pt]{article}

\usepackage{fullpage}
\usepackage{algorithm}
\usepackage{algorithmic}
\usepackage[colorlinks=true,linkcolor=black,citecolor=black,urlcolor=blue,pdfborder={0 0 0}]{hyperref}
\usepackage{graphicx}
\usepackage[caption=false]{subfig}
\usepackage{amssymb}
\usepackage{amsmath}
\usepackage{lscape}
\usepackage{booktabs}
\usepackage{natbib}
\usepackage{float}
\usepackage{tikz}

\newcommand{\transp}{^{^{\top}}}
\newcommand{\unit}[1]{\ensuremath{\, \mathrm{#1}}}
\newcommand{\bd}[1]{\dot{\textbf{#1}}} 

\begin{document}

\title{Mixed integer programming improves comprehensibility and plan quality in inverse optimization of prostate HDR-brachytherapy}

\author{Bram L. Gorissen$^1$, Dick den Hertog$^1$ and Aswin L. Hoffmann$^2$ \\ \\
\textit{\small $^1$ Department of Econometrics and Operations Research/Center for Economic Research} \\
\textit{\small (CentER), Tilburg University, PO Box 90153, Tilburg 5000 LE, the Netherlands,} \\
\textit{\small $^2$ Department of Radiation Oncology (MAASTRO), GROW School for Oncology and} \\
\textit{\small Developmental Biology, Maastricht University Medical Center, Maastricht 6201 BN,} \\
\textit{\small the Netherlands,} \\
\textit{\small {\tt b.l.gorissen@tilburguniversity.edu}}}
\date{}
\maketitle

\begin{abstract}
Current inverse treatment planning methods that optimize both catheter positions and dwell times in prostate HDR brachytherapy use surrogate linear or quadratic objective functions that have no direct interpretation in terms of dose-volume histogram (DVH) criteria, do not result in an optimum or have long solution times.

We decrease the solution time of existing linear and quadratic dose-based programming models (LP and QP, respectively) to allow optimizing over potential catheter positions using mixed integer programming. An additional average speed-up of 75\% can be obtained by stopping the solver at an early stage, without deterioration of the plan quality. For a fixed catheter configuration, the dwell time optimization model LP solves to optimality in less than 15 seconds, which confirms earlier results. We propose an iterative procedure for QP that allows to prescribe the target dose as an interval, while retaining independence between the solution time and the number of dose calculation points. This iterative procedure is comparable in speed to the LP model, and produces better plans than the non-iterative QP.

We formulate a new dose-volume based model that maximizes $V_{100\%}$ while satisfying pre-set DVH-criteria. This model optimizes both catheter positions and dwell times within a few minutes depending on prostate volume and number of catheters, optimizes dwell times within 35 seconds, and gives better DVH statistics than dose-based models. The solutions suggest that the correlation between objective value and clinical plan quality is weak in existing dose-based models.
\end{abstract}

\begin{tikzpicture}[remember picture,overlay]
\node[anchor=south,yshift=10pt] at (current page.south) {\fbox{\parbox{\dimexpr\textwidth-\fboxsep-\fboxrule\relax}{\footnotesize This is an author-created, un-copyedited version of an article published in Physics in Medicine and Biology \href{http://dx.doi.org/10.1088/0031-9155/58/4/1041}{DOI:10.1088/0031-9155/58/4/1041}.}}};
\end{tikzpicture}

\section{Introduction}
\subsection{HDR brachytherapy optimization}
Interstitial high-dose-rate (HDR) brachytherapy is a form of internal radiation therapy where a high activity iridium-192 stepping source is temporarily placed into the tumour volume or its proximity through the use of implanted catheters. This type of radiotherapy has shown to be an excellent option for the definitive treatment of localized prostate cancer in any risk category \citep{Yamada2012}. Clinical outcome data shows high tumour control and low toxicity rates because of the precision and control with which highly conformal optimized HDR treatment can be delivered. 

Treatment planning is one of the steps in the process to deliver the prescribed dose, and entails two design problems. Firstly, the number and spatial configuration of the catheters to be implanted has to be determined. Secondly, the spatio-temporal source stepping pattern within the implanted catheters needs to be calculated. The number and configuration of catheters depend on the prostate shape, volume, and regional anatomy. Often a template is used for the transperineal implantation of needles. As each catheter offers a range of potential stopping points (dwell positions) where the source can stay for a predefined time (dwell time), the design problem has many degrees of freedom. Hence, the problem formulation to design an HDR brachytherapy plan comes down to: 1) determine where to insert the catheters in the template, and 2) determine where and for how long to stop the source inside the catheters to achieve adequate coverage of the planning target volume (PTV) while sufficiently limiting dose to surrounding organs at risk (OAR). 

Computerized techniques for anatomy-based inverse treatment planning of HDR bra-chytherapy enable solving these problems by calculating catheter configurations and source dwell time distributions, based on mathematical optimization models. These models require the above problem formulation to be translated into a mathematical framework that describes how the dose distribution depends on the decision variables. Furthermore, it is mandatory to establish several levels of abstraction to assess the dosimetric quality of the resulting treatment plan. In this paper, we discriminate between three levels of abstraction. At the highest level, a dose penalty function is used to assign a penalty measure to a treatment plan. At the intermediate level, dose-volume-histogram (DVH) statistics are used to evaluate the dose and dose-volume characteristics of a given dose distribution. At the lowest level, the opinion of a human expert forms a subjective judgement of the three dimensional dose distribution. Ideally, lower values at the highest level correspond to better plans when evaluated at the lowest level. We briefly discuss these three levels, as they are relevant for the remainder of this paper.

\begin{figure}[ht]
	\centering
		\subfloat[]{ \includegraphics[scale=0.6]{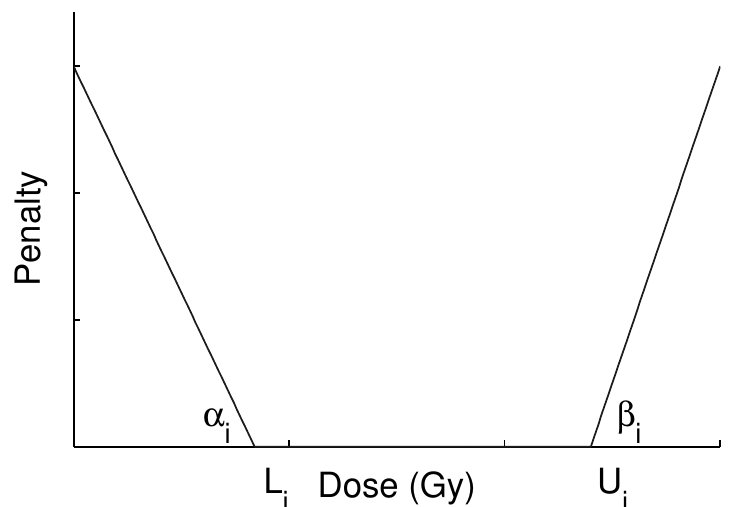} \label{fig:linpenalty} }
		\qquad
	  \subfloat[]{ \includegraphics[scale=0.6]{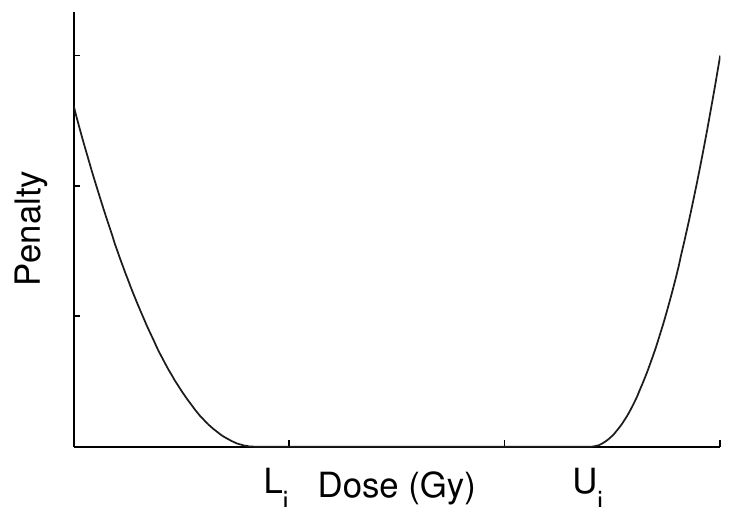} }
	\caption{The penalty in calculation point $i$ for not satisfying the lower or upper bound is either linear (a) or quadratic (b) in the violation.}
	\label{fig:penaltyfunctions} 
\end{figure}

\paragraph{Dose penalty functions.}
As it is impossible to calculate dose deposited to every single cell, the most relevant tissue volumes (i.e., prostate, urethra, rectum) are discretized into finite sets of dose calculation points that each represents an adequately small subvolume where the dose is considered to be uniform. At each point $i$, the delivered dose, $d_i$, is compared with a prescribed lower bound $L_i$ and upper bound $U_i$. If the delivered dose is not between these bounds, the violation is penalized. We use the linear and quadratic penalty function to develop our linear and quadratic dose-based optimization models, respectively (Figure \ref{fig:penaltyfunctions}). For a linear penalty function, costs of $\alpha_i$ or $\beta_i$ are incurred per unit dose (Gy) violation of the lower or upper bound, respectively, whereas for a quadratic penalty function a dose violation is penalized to the second degree. The total penalty per tissue structure is the summed penalty over all calculation points within the structure. 

The linear penalty function has been used in two well-known algorithms for anatomy-based inverse treatment planning: Inverse Planning by Simulated Annealing (IPSA) \citep{Lessard2001,Alterovitz2006} and Hybrid Inverse treatment Planning and Optimization (HIPO) \citep{Karabis2005}. The quadratic penalty function has also been discussed in the literature \citep{Milickovic2002,Lahanas2003,LahanasQ2003}.

\paragraph{DVH statistics.}
A typical example of clinically relevant dose-volume criteria for an HDR brachytherapy prostate plan with 2 fractions of 8.5 Gy following external beam radiation treatment with 13 fractions of 2.75 Gy is listed in Table \ref{tbl:clineval}. The criterion $D_{90\%} \geq 90\%$ requires the hottest $90\%$ of the PTV to receive at least 90\% of the prescribed dose, and $V_{150\%} \leq 55\%$ requires the relative volume that is exposed to more than $150\%$ of the prescribed dose to be less than $55\%$. For the urethra, $D_{0.1cc}$ $\leq 10 \unit{Gy}$ means that the minimum dose delivered to the hottest $0.1$ cc  does not exceed $10 \unit{Gy}$.

\begin{table}[ht]
	\caption{\label{tbl:clineval}Local protocol based on \protect\citep{Hoskin2007} of DVH criteria for a prescribed dose of 8.5 Gy per fraction.}
	\centering
		\begin{tabular}{lll}
			\toprule
			PTV & Rectum & Urethra \\
			\midrule
			$D_{90\%}  \geq 90\%$  &  $D_{10\%} \leq 7.2 \unit{Gy}$  &  $D_{10\%}  \leq 10 \unit{Gy}$ \\
			$V_{100\%} \geq 90\%$  &  $D_{2cc}  \leq 6.7 \unit{Gy}$  &  $D_{0.1cc} \leq 10 \unit{Gy}$ \\
			$V_{150\%} \leq 55\%$  &  $V_{94\%}    = 0   \unit{cc}$  &  $V_{125\%}  =    0 \unit{cc}$ \\
			$V_{200\%} \leq 20\%$  &  & \\
			\bottomrule
		\end{tabular}
\end{table}

\paragraph{Expert opinion.}
Clinically established DVH criteria are often used as quality indicators of a treatment plan. Computer optimized treatment plans often do not satisfy pre-set DVH criteria, and hence require {\it a posteriori}\ adjustment of the dwell time distributions. This is often accomplished using so-called `graphical optimization', see e.g. \citep{Morton2008}, or manual adaptation of individual dwell times. The quality of an optimization method can therefore also be expressed by the time spent to post-process the plan. It is obvious that the perceived plan quality strongly depends on the level of experience of the treatment planner.

\paragraph{Mixed integer programming.}
There is a matured field of research that deals with optimization problems where (some of the) variables have to be integer, which are inherently difficult to solve to optimality. When applied to problems in which variables are restricted to 0 or 1, the method starts by solving a relaxed problem where the binary variables are allowed to take any value between 0 and 1. If in the optimal solution there is at least one binary variable which is not integer, the method procedes by either (1) adding a constraint that does not exclude the optimal binary solution, but excludes the current optimal solution, or (2) splitting the problem into two subproblems, one with a selected binary variable fixed to 0 and one with the same binary variable fixed to 1. These steps may be combined, and are repeated until the optimal solution is found. This way, the method can report an upper and a lower bound on the objective value, where one corresponds to the current best solution and the other to the solution of a relaxed problem. We refer the interested reader to \citep{NemhauserWolsey1999}.

\subsection{Our contributions}
Currently, only HIPO can solve the problem of catheter placement for prostate HDR brachytherapy. However, because HIPO is partially based on heuristical optimization, mathematical optimality of  the solution cannot be guaranteed. This means that there may be solutions with lower objective values, that better satisfy the prescribed dose. We apply modifications to existing dose-based optimization models, so that a state-of-the-art mixed-integer linear programming (MILP) or mixed-integer quadratic programming (MIQP) solver can solve them to optimality.

Conventional quadratic dose-based optimization models have the advantage that the complexity of the optimization problem is independent of the number of dose calculation points if the target dose is specified as a single value rather than the interval $[L_i, U_i]$ \citep{Lahanas2003,LahanasQ2003}. This implies that underdosage and overdosage are penalized equally, even though the latter is allowed up to $U_i$. It is our second contribution to present an iterative procedure for quadratic penalty functions that retains the advantage of the conventional quadratic dose-based models.

As a third contribution, we present a new model that maximizes PTV coverage while constraining DVH parameters on the OAR(s). This model has a more direct clinical interpretation than linear penalty functions, solves in clinically acceptable time and is expected to produce better treatment plans than dose-based models. Using this model, we show that dose-based objective functions are bad surrogates for dose-volume based plan evaluation criteria.

The structure of this paper is as follows. We start by introducing the mathematical notations and optimization models in Section \ref{sec:methods}. Our methods are numerically evaluated on three clinical sample cases in Section \ref{sec:numeval}. The relation to other work is discussed in Section \ref{sec:discussion}.

\section{Methods}
\label{sec:methods}
\subsection{Mathematical notation}
\noindent We use the sets, parameters and variables listed in Tables \ref{tbl:sets}-\ref{tbl:vars}. The dose delivered to calculation point $i$ is a linear function of dwell time and dose rate:
\begin{eqnarray} 
	(\bd{D}\textbf{t})_i = \sum_{j\in J}\dot{d_{ij}}t_j. \label{dose-calcpoint}
\end{eqnarray}
\label{sec:notation} 
\begin{table}[H]\small
\vspace{-1cm}
	\caption{\label{tbl:sets}Sets used for inverse treatment planning.}
	\centering
		\begin{tabular}{ll}
			\toprule
			set & description \\
			\midrule
			$S$         & Tissue structures \{PTV, R(ectum), (U)rethra\} \\
			$I_s$       & Calculation points in structure $s \in S$\\ 
			$I$         & All calculation points, $I_{PTV} \cup I_R \cup I_U$ \\ 
			$K$         & Catheters \\
			$J_k$       & Dwell positions in catheter $k \in K$ \\ 
			$J$         & All dwell positions, $\bigcup_{k \in K}J_k$ \\ 
			$\Gamma(j)$ & Dwell positions adjacent to dwell position $j \in J_k$ within catheter $k$\\
			\bottomrule
		\end{tabular}
\end{table}
\begin{table}[H]\small
  \caption{\label{tbl:parameters}Parameters used for inverse treatment planning.}
	\centering
		\begin{tabular}{llp{10cm}}
			\toprule
			parameter     & unit       & description \\
			\midrule
			$N$            & 1          & Upper bound on the number of catheters allowed \\
			$t_{\mbox{\tiny max}}$     & s    & An upper bound on the dwell time for a single dwell position \\
			$\dot{d_{ij}}$ & Gy         & The dose rate delivered to calculation point $i \in I$ by a source at dwell position $j \in J$ per unit of time \\
			$\bd{D}$       & Gy         & The first order dose kernel matrix, i.e. the matrix with elements $\dot{d_{ij}}$ \\
			$L_i$          & Gy         & Prescribed lower bound on the dose for calculation point $i \in I$ \\
			$U_i$          & Gy         & Prescribed upper bound on the dose for calculation point $i \in I$ \\
			$p_i$          & Gy         & Prescribed dose for calculation point $i \in I$ (in case $L_i=U_i$) \\
			$\textbf{p}$       & Gy         & The prescribed dose vector with elements $p_i$ \\
			$\tau_s$       & 1          & Percentage of calculation points receiving a dose less than $L_i$ in structure $s \in S$ \\
			$\alpha_i$     & Gy$^{-1}$  & Penalty per Gy below the lower bound $L_i$ for calculation point $i \in I$ \\
			$\beta_i$      & Gy$^{-1}$  & Penalty per Gy exceeding the upper bound $U_i$ for calculation point $i \in I$ \\
			$\gamma$       & \%         & The maximum allowable relative difference in dwell times between two adjacent dwell positions \\
			$w_i$          & 1          & Relative weight of calculation point $i$, $w_i = 1/|I_s|$ where $s$ is the structure containing $i$ \\
			$\textbf{W}$       & 1          & Weight matrix, with $w_i$ on the diagonal and 0 at other positions \\
			\bottomrule
		\end{tabular}
\end{table}
\begin{table}[H]\small
	\centering
  \caption{\label{tbl:vars}Variables used for inverse treatment planning.}
		\begin{tabular}{llp{10cm}}
			\toprule
			variable & unit & description \\
			\midrule
			$t_j$    & s    & Dwell time at dwell position $j \in J$ \\ 
			$\textbf{t}$ & s    & The dwell time vector with elements $t_j, j \in J$ \\
			$v_i$    & 1    & Binary variable indicating whether calculation point $i$ receives at most (or at least) its prescribed dose \\ 
			$b_k$    & 1    & Binary variable indicating whether catheter position $k \in K$ in the treatment template is used \\ 
			$x_i$    & 1    & Penalty for calculation point $i \in I$ \\ 
			\bottomrule
		\end{tabular}
\end{table}
\subsection{Optimization models}
\label{sec:models}

\subsubsection{Linear Dose-based (LD) model.}
\label{sec:linmodel}
Using the notation introduced in Section \ref{sec:notation}, the linear dose-based objective function in Figure \ref{fig:linpenalty} can be written as:
\begin{eqnarray}
	\min  	\qquad	&\sum_{i\in I} \max \{0, \alpha_i (L_i - \sum_{j\in J}\dot{d_{ij}}t_j), \beta_i (\sum_{j\in J}\dot{d_{ij}}t_j - U_i)\}. \label{obj}
\end{eqnarray}

\noindent In order to transform this objective function into a model with a linear objective function and linear constraints, we introduce a variable $x_i$ replacing the argument of the max operator in \eqref{obj}. Additional constraints limit the number of catheters used and the relative difference in dwell time between adjacent positions. The latter implements the dwell time modulation restriction (DTMR), that is often used to prevent hot spots \citep{Baltas2009}. The full model then becomes:
\begin{subequations}
\begin{eqnarray}
	(LD) \quad	\min  \quad    &\sum_{i\in I} w_i x_i                      & \label{lin-obj} \\
	\mbox{s.t.}  &x_i \geq \alpha_i [ L_i - \sum_{j\in J}\dot{d_{ij}}t_j] \quad  & \forall i\in I \label{lin-robustify1} \\
               &x_i \geq \beta_i [ \sum_{j\in J}\dot{d_{ij}}t_j - U_i]         & \forall i\in I \label{lin-robustify2} \\
               &t_j \leq b_k t_{\mbox{\tiny max}}                        & \forall k \in K \quad \forall j\in J_k \label{lin-activate} \\
               &t_{j_{1}} \leq (1+100\gamma) t_{j_{2}}                   & \forall j_{1}\in J \quad \forall j_{2}\in \Gamma(j_{1}) \label{lin-modul} \\
               &\sum_{k\in K}b_k \leq N                                  & \label{lin-numcath} \\
               &b_k \in \{0,1\}                                          & \forall k\in K \\
               &x_i \geq 0                                               & \forall i\in I \label{lin-nonnegx} \\
               &t_j \geq 0                                               & \forall j\in J. \label{lin-nonnegt}
\end{eqnarray}
\end{subequations}

The penalty function for calculation point $i$ is a convex piecewise linear function in the dose $\sum_{j\in J}\dot{d_{ij}}t_j$. Constraints \eqref{lin-robustify1}, \eqref{lin-robustify2} and \eqref{lin-nonnegx} together with objective \eqref{lin-obj} make $x_i$ equal to the pointwise maximum. If catheter $k$ is not used then $b_k = 0$. Constraint \eqref{lin-activate} sets the dwell times within that catheter to 0 seconds. Constraints \eqref{lin-modul} and \eqref{lin-numcath} implement the DMTR and enforce no more than $N$ catheters to be selected, respectively.

If the DTMR would be dropped, the model is equivalent to the one described by \citet{Karabis2009}. If additionally the values of $b_k$ are fixed, i.e., if all catheter positions are fixed, the model corresponds to the one by \citet{Alterovitz2006}.

When we solve it as an MILP, the solution times are very high and clinically unacceptable. This is in line with the results from \citet{Karabis2009}, where a similar model could not be solved in less than $5$ hours when the number of catheter positions was more than 25--30.

We have improved the solution time by making two improvements. The first is to specify constraint \eqref{lin-activate} as an indicator constraint, which is an option offered by our solver that helps treating this constraint more efficiently. Only if $b_k = 0$, the constraint $t_j=0$ becomes visible to the solver. The second improvement is to make two adjacent catheter positions mutually exclusive by adding an exclusion restriction $b_{k_1}+b_{k_2}\leq 1$ for any two catheters $k_1$ and $k_2$ that are adjacent in the template. The rationale for this is that two adjacent catheters are likely to cause high-dose subregions to become connected and form undesirable hot spots.

\subsubsection{Quadratic Dose-based (QD) model.}
\label{sec:quadratic}
As an alternative to the (LD) model, we propose a convex quadratic model. If we use a quadratic objective function, the number of calculation points no longer plays a role, thus greatly reduces complexity. By using identity (\ref{dose-calcpoint}), $(\bd{D}\textbf{t})_i$ it is evident that $\sum_i(w_i (\textbf{D}\textbf{t})_i-p_i)^{2}$ measures the deviation from the prescribed dose $p_i$ in calculation point $i$. This can also be written as the squared 2-norm $||\textbf{W}(\bd{D}\textbf{t}-\textbf{p})||_2^2 $, which is convex in $\textbf{t}$. Consider the following constrained least-squares approximation model:
\begin{subequations}
\begin{eqnarray}
	(QD) \qquad \min \quad &|| \textbf{W}(\bd{D}\textbf{t}-\textbf{p}) ||_{2}^{2} \quad & \\
	\mbox{s.t.} \quad &t_j\leq b_k t_{\mbox{\tiny max}}      & \forall k \in K \quad \forall j\in J_k \\
                  &t_{j_{1}}\leq (1+100\gamma) t_{j_{2}}   & \forall j_{1}\in J \quad \forall j_{2}\in \Gamma(j_{1}) \\
                  &\sum_{k\in K}b_k\leq N                  & \\
                  &b_k\in \{0,1\}                          & \forall k\in K \\
                  &t_j \geq 0.                             &
\end{eqnarray}
\end{subequations}

The objective can be rewritten as: $(\textbf{W}(\bd{D}\textbf{t}-\textbf{p}))\transp (\textbf{W}(\bd{D}\textbf{t}-\textbf{p})) = \textbf{t}\transp \bd{D}\transp \textbf{W} \transp \textbf{W} \bd{D}\textbf{t}- 2 \textbf{p}\transp \textbf{W}\transp \textbf{W} \bd{D}\textbf{t} + \textbf{p} \transp \textbf{W} \transp \textbf{W} \textbf{p}$. Instead of the full $|I| \times |J|$ matrix $\bd{D}$ it suffices to specify the $|J| \times |J|$ matrix $\bd{D}\transp \textbf{W} \transp \textbf{W} \bd{D}$ and the $|J| \times |1|$ vector $\bd{D}\transp \textbf{W}\transp \textbf{W} \textbf{p}$, whose sizes do not increase with the number of calculation points. The latter has been observed before by \citet{Lahanas2003}, but has not been used for formulating a convex quadratic programming model. Instead, these authors formulate nonconvex models, for which optimality cannot be guaranteed. However, the (QD) model can be solved to optimality. The solution time is greatly reduced by adding the exclusion restriction. 

In this model, the prescribed dose $p_i$ for calculation point $i$ has to be specified. It is not immediately clear which value should be taken. For points outside the PTV, a value of $0$ is reasonable. For points inside the PTV it is difficult not to penalize very reasonable values. Finding a good value for $p_i$ only gives a target value that is good for the average calculation point. All calculation points will still contribute some amount to the objective function even though they receive a dose between $L_i$ and $U_i$. We can alleviate this disadvantage by solving the problem iteratively. The algorithm starts by initializing each $p_i$ at $\left(L_i + U_i\right)/2$. For each iteration, first the problem is solved, then $p_i$ gets adjusted to a value in $[L_i, U_i]$ closest to the dose received in the current optimal solution. The algorithm stops when the improvement in objective value is sufficiently small. A very precise description is given in Algorithm \ref{alg:iterative} in \ref{ap:alg}. This procedure in general does not necessarily converge to the global optimum that could have been obtained by minimizing simultaneously over $\textbf{t}$, $\dot{d_{ij}}$ and $b_k$, which is proven with a small example in \ref{sec:counterexample}.

\subsubsection{Linear Dose-Volume based (LDV) model.}
\label{sec:dvh}
We propose a new model that maximizes the fraction of the PTV receiving the prescribed dose, while constraining DVH parameters for OARs. For the rectum and urethra we enforce $D_{10\%} \leq 7.2 \unit{Gy}$ and $D_{10\%} \leq 10 \unit{Gy}$, respectively. In accordance with Table \ref{tbl:clineval}, we do not allow a dose higher than $8 \unit{Gy}$ in the rectum, and $10.6 \unit{Gy}$ in the urethra. This formulation has the advantage that a feasible solution exists (e.g. take all dwell times equal to $0$), and that the solution shows the best target coverage that satisfies clinically derived DVH constraints. We formulate the model as:
\begin{subequations}
\begin{eqnarray} 
	(LDV) \quad \max  \quad    & \frac{1}{|I_{PTV}|} \sum_{i\in I_{PTV}} v_i & \label{bin-obj} \\
	\mbox{s.t.} \quad  &\sum_{j\in J}\dot{d_{ij}}t_j \geq L_i v_i      \quad  & \forall i \in I_{PTV}       \label{bin-V100} \\
               &\sum_{j\in J}\dot{d_{ij}}t_j \leq L_i + (U_i-L_i)(1-v_i)    & \forall i \in I_R \cup I_U  \label{bin-D1} \\
               &\sum_{i\in I_{s}} v_i \geq \tau_s |I_{s}|             & \forall s \in \{ R,U \}     \label{bin-D2} \\
               &t_j \leq b_k t_{\mbox{\tiny max}}                     & \forall k \in K \quad \forall j\in J_k                   \label{bin-activate} \\
               &t_{j_{1}} \leq (1+100\gamma) t_{j_{2}}                & \forall j_{1} \in J \quad \forall j_{2}\in \Gamma(j_{1}) \label{bin-modul} \\
               &\sum_{k\in K}b_k \leq N                               &                  \label{bin-numcath} \\
               &b_k \in \{0,1\}                                       & \forall k \in K  \label{bin-bincath} \\
               &v_i \in \{0,1\}                                       & \forall i \in I  \label{bin-binclin} \\
               &t_j \geq 0                                            & \forall j \in J. \label{bin-nonnegt}
\end{eqnarray}
\end{subequations}
Here $L_i$ and $U_i$ have a slightly different interpretation than in the dose-based models. We select $L_i = 8.5 \unit {Gy}$ for the PTV, $(L_i,U_i) = (7.2,8) \unit {Gy}$ for the rectum, $(L_i,U_i) = (10,10.6) \unit {Gy}$ for the urethra and $\tau_R = \tau_U = 0.9$.

Constraint (\ref{bin-V100}) allows $v_i$ for $i \in I_{PTV}$ to be $1$ only if the dose exceeds $L_i$. Hence objective function (\ref{bin-obj}) maximizes the number of points inside the PTV that receive the prescribed dose. Constraint (\ref{bin-D1}) allows $v_i$ for $i \in I_{R}$ to be $1$ only if the dose is less than $7.2 \unit{Gy}$, and never allows a dose higher than $8 \unit{Gy}$. Constraint (\ref{bin-D2}) then enforces 90\% of the $v_i$ to be $1$. Similarly, the same constraints enforce that 90\% of the urethra receives a dose less than $10 \unit{Gy}$, and no part in the urethra receives a dose higher than $125\%$. Constraints (\ref{bin-activate})--(\ref{bin-numcath}) are the same as constraints (\ref{lin-activate})--(\ref{lin-numcath}) in the (LD) model.

By fixing the parameters $b_k$, (LDV) becomes a dwell time optimization model. This enables a comparison with IPSA, which is not able to optimize catheter positions.

One of the advantages of directly optimizing on clinically relevant criteria is the possibility to extend the model in a clinically interpretable way. Suppose for instance that the maximum number of catheters is not fixed {\it a priori}, but the planner wants to insert an extra catheter only if it leads to an improvement of $V_{100\%}$ of at least 5\%. This can be incorporated into (LDV) by changing the objective to $1/|I_{PTV}| \sum_{i\in I_{PTV}} v_i + 0.05N$ and by treating $N$ as a variable rather than as a parameter.
\section{Numerical evaluation}
\label{sec:numeval}

\subsection{Patient data}
Clinical data from three different patients have been obtained from the treatment planning system (HDRplus, version 3.0, Eckert \& Ziegler BEBIG GmbH, Berlin, Germany). Characteristics are summarized in Table \ref{tbl:patientdata}. Approximately 2500 calculation points have been hexagonally distributed over the PTV, rectum and urethra. The number of potential catheter positions in the template is 40, 49 and 43 for patient 1, 2 and 3, having a prostate volume of 50 cc, 75 cc and 81 cc, respectively. According to our clinical protocol, the PTV had been extended with a 2 mm margin, and dwell positions were activated with a separation of 3 mm. A transperineal needle template with a hole resolution of 5 mm was used (Martinez Prostate Template, Nucletron BV, Veenendaal, the Netherlands). The dose rates have been calculated using the TG-43 formalism with the source parameters according to \citep{Granero2006}.

\begin{table}[ht]\small
	\caption{\label{tbl:patientdata}Characteristics of the patient data.}
	\centering
		\begin{tabular}{lllllllll}
			\toprule
			                   &            &       &            &       & \multicolumn{3}{c}{Number of calculation points} \\
\cmidrule(lr){6-8} 
			\textrm{Structure} & $\alpha_i$ & $L_i$ & $\beta _i$ & $U_i$ & \textrm{Patient 1} & \textrm{Patient 2} & \textrm{Patient 3}\\
			\midrule
			PTV                & $8$ & $8.5$ & $3$ & $25$    & $1732$  & $1834$   & $1791$ \\
			Rectum             & $0$ & $0$   & $10$  & $8$   & $246$ & $234$  & $240$ \\
			Urethra            & $0$ & $0$   & $10$  & $10$    & $489$ & $473$  & $495$ \\
			\bottomrule
		\end{tabular}

\end{table}

\subsection{Inverse planning simulated annealing (IPSA)}
We compare our results with the IPSA implementation in HDRplus, which exploits the linear penalty function \eqref{obj} and was configured as follows. The composite objective function did not include the total dwell time. A maximum weight was used for the DTMR. The trade-off between speed and quality was set to its default value. After three consecutive runs, the plan with the lowest objective value was selected.

\subsection{Our optimization models}

For the model parameter values we set $t_{\mbox{\tiny max}}$ at $5$ seconds for an apparent source activity of 370 GBq. The DTMR parameter $\gamma$ was set at $10\%$, and the maximum allowed number of catheters $N$ was varied between $15$ and $20$.

All computing times reported have been obtained with the optimization software AIMMS 3.10 x64 using ILOG CPLEX 12.1 as solver running on Windows 7 x64 on an Intel Core i5 660 (3.33 GHz) processor with 8 GB of RAM.

The exclusion restriction reduces the number of allowed catheter configurations with a factor $10^3$--$10^8$, depending on the prostate volume and number of catheters.

The solution time and objective values of all models are listed in Tables \ref{tbl:numeval1}--\ref{tbl:numeval3}. The first line is read as follows: for patient 1, the IPSA model, which is optimized for $16$ preselected catheters, returns an optimized plan within $0.8$ minutes. If the (LD) model, (QD) model, Algorithm 1 or the (LDV) model had chosen the same dwell times as IPSA, they would have given objective values of $0.9280$, $189$, $1.1229$ or $86.1$, respectively. The other columns show the dosimetric plan performance.

\subsubsection{Linear dose-based optimization}
The solution times for the (LD) model are 5, 364 and 3 minutes for patients 1, 2 and 3, respectively when the allowed number of catheters is $20$. The high solution time for patient 2 is probably due to the large number of feasible catheter configurations. We have not been able to obtain a solution within $24$ hours without the exclusion restriction. Specifying constraint \eqref{lin-activate} as an indicator constraint decreases calculation time by about 10\%.

\begin{figure}
	\includegraphics[scale=0.6]{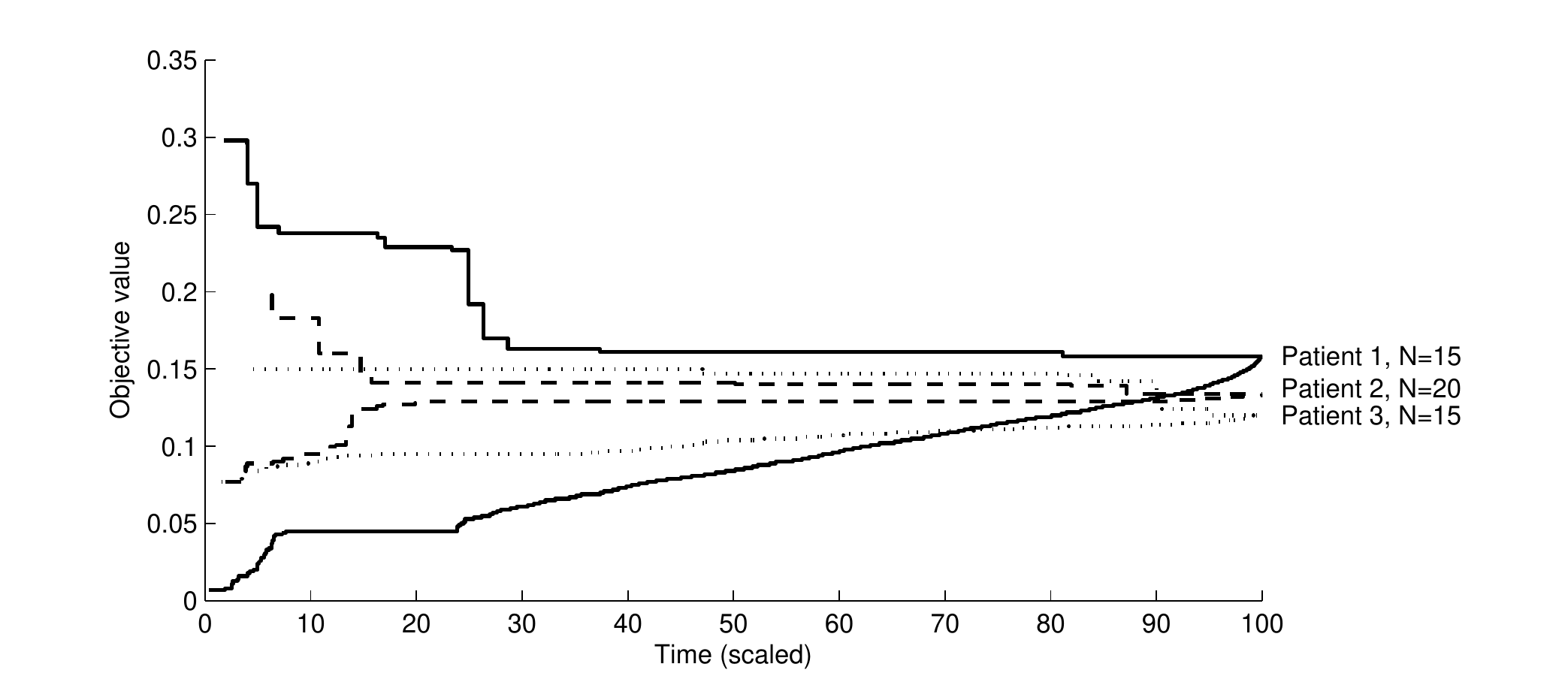}
	\caption{Convergence of the lower and upper bound during the optimization of the (LD) model.}
	\label{fig:mip-progress}
\end{figure}

The convergence rate of the lower and upper bound of $V_{100\%}$ during the optimization process is depicted in Figure \ref{fig:mip-progress}. We observe that most time is spent on obtaining a better lower bound. If we would terminate the solver when the upper bound is at most twice the lower bound, we would get a solution four times quicker at the cost of a 5\% higher objective value. A slightly higher objective value does not translate into a significantly lower plan quality when evaluated at a lower level \citep{Alterovitz2006,Karabis2009}. Hence, stopping early will on average not result in clinically worse treatment plans.

When the catheter positions are fixed, the (LD) model reduces to an LP. CPLEX finds the optimal solution in $2.4$ seconds. This is comparable to the solution time reported for a similar LP \citep{Alterovitz2006,Karabis2009}.

\subsubsection{Quadratic dose-based optimization}
First, we searched for a good value of $p_i$ in the PTV. By repeatedly solving the model for different target doses and evaluating the dose distribution, we found that 21 Gy gave an acceptable treatment plan for patient 1. We used the same target value for patient 2 and 3.

The (QD) model solved in between $18$ seconds and $1.5$ minutes for $N=20$ catheter positions. Again, we observed that the solution time decreased with a larger allowed number of catheters. Comparing the solution times to those of the (LD) model, we found that the (QD) model was at least ten times faster.

For the iterative procedure (Algorithm \ref{alg:iterative}), we observed a drop in objective value to $5\%$ of the initial value after the first iteration. In the subsequent $15$ steps, the additional decrement was 40--50\%. After 14--24 iterations in total, the decrements in objective value were smaller than $10^{-3}$, which is when the procedure was halted.

When the catheter positions are fixed, and only the dwell times are to be optimized, the (QD) model solves in 0.14, 0.9 and 0.9 seconds for patients 1, 2 and 3 respectively, while Algorithm \ref{alg:iterative} solves in 8.5, 13.3, and 11.8 seconds, respectively.

\subsubsection{Dose-volume based optimization}
The solver requires more than 24 hours to solve the (LDV) model to optimality. This is not problematic, because during execution the solver reports both $V_{100\%}$ of the best solution found so far, and an upper bound on $V_{100\%}$ that gradually gets lower. Hence, in a clinical setting, the treatment planner can stop the solver as soon as the value of $V_{100\%}$ is satisfactory. Here, we stopped as soon as $V_{100\%} \geq 95\%$ or after $15$ minutes. Stopping the solver before optimality is reached is not a novel idea. It has been applied before with low-dose-rate brachytherapy \citep{Gallagher1997,Lee2003}. For dwell time optimization, the solution time is 8.5 minutes for patient 2. By changing solver parameters, we have reduced the solution time for dwell time optimization to 14, 30 and 35 seconds for patients 1, 2 and 3, respectively. The new parameters make the solver first consider solutions in which binary variables that are close to 0 in the LP relaxation, are fixed at 0. At least 90\% of the binary variables must be 1. After fixing 10\% of those variables to 0, the other variables must take the value 1 in the LP relaxation \citep{Pryor2011}. We have also tried Benders' decomposition \citep{Benders1962}, but it did not provide a speed-up.

\subsection{Plan performance evaluation} All models generate plans with good OAR sparing. The DVH bounds on the rectum and the urethra are (almost) satisfied in all cases. The $V_{100\%}$ for the PTV is sufficiently high for all models except the (QD) model. $V_{150\%}$ and $V_{200\%}$ are sufficiently low for all models, except for the (QD) model. 

The plan performance as assessed from the three-dimensional (3D) dose distribution by an experienced planner (ALH) is listed in the last column of Tables \ref{tbl:numeval1}-\ref{tbl:numeval3}. The experienced planner paid most attention to conformality of the dose distribution and to whether high dose subvolumes (150\% and 200\%) around dwell positions were connected. Indeed a high correlation was found between the plan performance scored by the expert and the COIN value.

The (QD) model produces inacceptable plans due to a low $V_{100\%}$, especially for patient 2. All other models perform very well, the (LDV) model for patient 1 being an exception. The latter is due to the activation of dwell positions outside the PTV, giving rise to significant dose contributions outside the PTV. This can probably be avoided by activating dwell positions only inside the PTV or by adding constraints on auxiliary avoidance structures that limit dose outside the PTV.

The relation between the linear penalty function value and the expert opinion is very weak. This becomes clear from the objective values of the solution of the (LDV) model: for patient 3 with 18 catheters the plan of the (LDV) model is still preferred by the expert even though the linear objective value is 4.6 times higher (0.1038 vs.~0.4816) than the optimal plan of the (LD) model. Also the relation between the linear objective value and DVH statistics is weak. For patient 1 with 16 catheters, the plans of the (LD) and (LDV) models have similar DVH statistics, but their linear penalty value differs by a factor 12 (0.1338 vs. 1.6512).

The effect on the clinical evaluation criteria of extra catheters above 15 is small for all three patients. In most cases, the models could slightly improve $V_{100\%}$ by increasing the number of catheters, the largest improvement being 2.1\%. The expert opinion is almost constant for a specific patient and model, with most variability related to patient 3. For this patient, the expert opinion always becomes more positive when more catheters were inserted.
\section{Discussion}
\label{sec:discussion}
With existing dose-based models for inverse treatment planning of HDR brachytherapy, it is often a trial-and-error process to obtain adequate treatment plans that satisfy pre-set DVH criteria. In a clinical setting, this is undesirable because of the time burden and the required degree of user experience. Often, the aim is to design a plan with maximum achievable target coverage under fixed OAR dose-volume constraints. This type of optimization problem can be formulated and solved using mixed integer programming. This improves comprehensibility and plan quality compared to traditional dose-based inverse optimization. We investigated enhancements of existing linear and quadratic programming models for dose-based optimization and showed that the solution time could be decreased substantially.

We have limited our analysis to three representative clinical cases that cover a range of prostate sizes. The limited number of patients has allowed us to perform an extensive analysis of the effects of different algorithms on the allowed number of catheters. We realize that it is necessary to include more patients to further strengthen our conclusions.

In this dosimetric study, perturbations unavoidable in a clinical implementation were not taken into account. There is still a lack of validated data on the uncertainties, which makes it hard to assess the impact of these uncertainties on the optimality of a treatment plan. In future work the uncertainties need to be identified and quantified, and the effect on treatment plans needs to be evaluated. Current optimization techniques that deal with uncertainties such as ``stochastic programming'' or ``robust optimization'' require a model that finds good treatment plans when there is no uncertainty \citep{BenTal,Kall}. This means that our work is also relevant for a future study on finding robust treatment plans.

For dwell time optimization, all models except (QD) can produce clinically good plans. This confirms current practice where the (LD) model is used.

Choosing catheter positions is still a difficult problem. Despite the exclusion restriction, only the (LDV) model produces plans in clinically acceptable time for all patients. There is one other article that uses an exclusion restriction, allowing a maximum of two catheters in any $2\times2$ square of template holes \citep{Holm2011}. This is weaker than our restriction. The HIPO algorithm is widely used and can optimize catheter positions in clinically acceptable time. HIPO inherently differs from our mixed integer programming approach, making it difficult to clarify any discrepancies.

DVH-based optimization has been applied to HDR brachytherapy by others \citep{Lahanas2003DVH,Panchal2008,Siauw2011}. All restrict to dwell time optimization and use heuristics for which mathematical optimality cannot be guaranteed. Of these, only \citet{Siauw2011} provide a fast solution to a MILP formulation, but cannot provide a good upper bound for $V_{100\%}$.

Weak correlation between the (LD) objective value and DVH parameters was also observed by \citet{Holm2011}. Holm claims that the (LD) objective can make a rough division between good and bad plans. We confirm that there could be an order of magnitude difference in (LD) objective values among good plans. This implies there is a gap between the objective function in dose-based models and clinically desired properties of a dose distribution. The (LDV) model has the potential to close this gap, and to give the planner a better tool to steer the optimization.

\section{Conclusion}
With the proposed extensions, existing dose-based optimization models that simultaneously optimize catheter positions and dwell times can be solved more quickly to proven optimality with mixed integer programming techniques. Our dose-volume based model relates more closely with clinical parameters compared to dose-based models, and is faster to solve.

\section*{Acknowledgments}
We thank Ulrich Wimmert$^\dagger$ from SonoTECH GmbH (Neu-Ulm, Germany) for providing a research version of HDRplus software that has the ability to export the dose rate kernel matrix and import the dwell times. We thank Frank Verhaegen from MAASTRO Clinic (Maastricht, the Netherlands) for his valuable comments and suggestions to improve this paper.

\bibliographystyle{abbrvnatnew}
\bibliography{personal}

\appendix

\section{Numerical evaluation}\label{ap-numeval}
\begin{landscape}
\begin{table}[T]\small{\center
\caption{\label{tbl:numeval1}Treatment plan performance indicators for patient 1.}
\begin{tabular}{llr llll llll ll ll l}
\toprule
  & & & & & & & \multicolumn{4}{c}{PTV (\%)}     & \multicolumn{2}{c}{Rectum (\unit{Gy})} & \multicolumn{2}{c}{Urethra  (\unit{Gy})} & \\
\cmidrule{8-11} \cmidrule{12-13} \cmidrule{14-15}
Model & $N$ & Sol. time & (LD) & (QD) & Alg. 1 & COIN & $D_{90\%}$ & $V_{100\%}$ & $V_{150\%}$ & $V_{200\%}$ & $D_{10\%}$ & $D_{2cc}$ & $D_{10\%}$ & $D_{0.1cc}$ & Expert \\
   & & (min) & value & value & value & & $\geq 90\%$ & $\geq 90\%$ & $\leq 55\%$ & $\leq 20\%$ & $\leq 7.2\unit{Gy}$ & $\leq 6.7\unit{Gy}$ & $\leq 10\unit{Gy}$ & $\leq 10\unit{Gy}$ & opinion \\
\midrule
IPSA$^\ast$ & 16 & $2.1$ & $0.109$ & $166$ & $0.029$ & $0.666$ & $110.8$ & $97.2$ & $46.6$ & $17.2$ & $6.5$ & $5.5$ & $9.6$ & $9.7$ & + \\
\midrule
{\bf (LD)} & {\bf 15} & $\mathbf{11.0}$  & $\mathbf{0.158}$ & $\mathbf{163}$ & $\mathbf{0.123}$ & $\mathbf{0.729}$ & $\mathbf{111.3}$ & $\mathbf{96.6}$ & $\mathbf{44.8}$ & $\mathbf{13.4}$ & $\mathbf{6.0}$ & $\mathbf{4.9}$ & $\mathbf{9.7}$ & $\mathbf{9.9}$ & {\bf ++} \\
{\bf (LD)} & {\bf 16} & $\mathbf{7.0}$ & $\mathbf{0.134}$ & $\mathbf{162}$ & $\mathbf{0.069}$ & $\mathbf{0.727}$ & $\mathbf{111.8}$ & $\mathbf{97.1}$ & $\mathbf{45.6}$ & $\mathbf{13.6}$ & $\mathbf{6.0}$ & $\mathbf{4.9}$ & $\mathbf{9.7}$ & $\mathbf{9.9}$ & {\bf ++} \\
{\bf (LD)} & {\bf 17} & $\mathbf{4.9}$ & $\mathbf{0.131}$ & $\mathbf{162}$ & $\mathbf{0.064}$ & $\mathbf{0.729}$ & $\mathbf{111.7}$ & $\mathbf{97.1}$ & $\mathbf{44.8}$ & $\mathbf{13.1}$ & $\mathbf{5.9}$ & $\mathbf{4.8}$ & $\mathbf{9.7}$ & $\mathbf{9.9}$ & {\bf ++} \\
{\bf (LD)} & {\bf 18} & $\mathbf{4.9}$ & $\mathbf{0.123}$ & $\mathbf{162}$ & $\mathbf{0.031}$ & $\mathbf{0.738}$ & $\mathbf{111.6}$ & $\mathbf{97.4}$ & $\mathbf{44.0}$ & $\mathbf{12.9}$ & $\mathbf{5.7}$ & $\mathbf{4.7}$ & $\mathbf{9.7}$ & $\mathbf{9.9}$ & {\bf ++} \\
{\bf (LD)} & {\bf 19} & $\mathbf{5.3}$ & $\mathbf{0.121}$ & $\mathbf{162}$ & $\mathbf{0.029}$ & $\mathbf{0.741}$ & $\mathbf{111.6}$ & $\mathbf{97.1}$ & $\mathbf{42.6}$ & $\mathbf{12.2}$ & $\mathbf{5.6}$ & $\mathbf{4.7}$ & $\mathbf{9.7}$ & $\mathbf{9.9}$ & {\bf ++} \\
{\bf (LD)} & {\bf 20} & $\mathbf{4.9}$ & $\mathbf{0.118}$ & $\mathbf{162}$ & $\mathbf{0.028}$ & $\mathbf{0.739}$ & $\mathbf{112.4}$ & $\mathbf{97.2}$ & $\mathbf{43.1}$ & $\mathbf{12.0}$ & $\mathbf{5.6}$ & $\mathbf{4.7}$ & $\mathbf{9.8}$ & $\mathbf{9.9}$ & {\bf ++} \\
\midrule
(QD) & 15 & $1.0$ & $1.296$ & $151$ & $1.417$ & $0.625$ & $99.3$ & $89.5$ & $50.0$ & $22.5$ & $5.7$ & $4.7$ & $8.8$ & $8.9$ & - \\
(QD) & 16 & $0.5$ & $1.238$ & $150$ & $1.101$ & $0.641$ & $98.7$ & $88.8$ & $50.1$ & $25.2$ & $5.5$ & $4.6$ & $8.9$ & $9.1$ & - \\
(QD) & 17 & $0.3$ & $1.167$ & $149$ & $1.015$ & $0.645$ & $99.0$ & $89.3$ & $52.0$ & $24.5$ & $5.5$ & $4.6$ & $9.1$ & $9.3$ & - \\
(QD) & 18 & $0.3$ & $1.167$ & $149$ & $1.015$ & $0.645$ & $99.0$ & $89.3$ & $52.0$ & $24.5$ & $5.5$ & $4.6$ & $9.1$ & $9.3$ & - \\
(QD) & 19 & $0.3$ & $1.167$ & $149$ & $1.015$ & $0.645$ & $99.0$ & $89.3$ & $52.0$ & $24.5$ & $5.5$ & $4.6$ & $9.1$ & $9.3$ & - \\
(QD) & 20 & $0.3$ & $1.167$ & $149$ & $1.015$ & $0.645$ & $99.0$ & $89.3$ & $52.0$ & $24.5$ & $5.5$ & $4.6$ & $9.1$ & $9.3$ & - \\
\midrule
Alg. 1 & 15 & $4.9$ & $0.341$ & $159$ & $0.055$ & $0.700$ & $106.4$ & $94.2$ & $46.5$ & $14.7$ & $5.8$ & $4.7$ & $9.4$ & $9.6$ & + \\
Alg. 1 & 16 & $4.0$ & $0.226$ & $160$ & $0.031$ & $0.713$ & $108.9$ & $95.6$ & $45.0$ & $13.7$ & $5.9$ & $4.8$ & $9.5$ & $9.7$ & + \\
Alg. 1 & 17 & $3.6$ & $0.243$ & $158$ & $0.032$ & $0.710$ & $106.9$ & $94.7$ & $45.2$ & $15.1$ & $5.7$ & $4.7$ & $9.4$ & $9.5$ & + \\
{\bf Alg. 1} & {\bf 18} & $\mathbf{3.2}$ & $\mathbf{0.218}$ & $\mathbf{158}$ & $\mathbf{0.025}$ & $\mathbf{0.715}$ & $\mathbf{109.7}$ & $\mathbf{96.1}$ & $\mathbf{47.0}$ & $\mathbf{14.8}$ & $\mathbf{5.7}$ & $\mathbf{4.7}$ & $\mathbf{9.7}$ & $\mathbf{9.9}$ & {\bf ++} \\
Alg. 1 & 19 & $2.8$ & $0.250$ & $158$ & $0.025$ & $0.714$ & $110.1$ & $96.2$ & $49.2$ & $14.7$ & $5.7$ & $4.7$ & $9.8$ & $10.0$ & + \\
Alg. 1 & 20 & $2.6$ & $0.245$ & $158$ & $0.023$ & $0.712$ & $110.5$ & $96.3$ & $51.1$ & $14.3$ & $5.8$ & $4.8$ & $9.8$ & $10.0$ & + \\
\midrule
(LDV) & 12 & $0.3$ & $3.002$ & $188$ & $22.249$ & $0.613$ & $113.6$ & $98.9$ & $52.3$ & $27.0$ & $6.2$ & $5.3$ & $9.8$ & $9.9$ & - - \\
(LDV) & 16 & $0.3$ & $1.651$ & $176$ & $8.558$ & $0.687$ & $113.4$ & $98.9$ & $43.2$ & $19.0$ & $6.0$ & $5.0$ & $9.9$ & $10.0$ & - - \\
{\bf (LDV)}$^\ast$ & {\bf 16} & $\mathbf{0.1}$ & $\mathbf{1.623}$ & $\mathbf{173}$ & $\mathbf{8.813}$ & $\mathbf{0.703}$ & $\mathbf{113.1}$ & $\mathbf{98.6}$ & $\mathbf{48.2}$ & $\mathbf{22.7}$ & $\mathbf{6.0}$ & $\mathbf{5.1}$ & $\mathbf{9.9}$ & $\mathbf{10.0}$ & {\bf ++}
\\
\bottomrule
\end{tabular}
\\
}  \scriptsize{$^\ast$ Using the catheter configuration chosen by an expert planner (ALH).} \\
  \mbox{Abbreviations: $N =$ allowed number of catheters, Sol.~time $=$ solution time, COIN $=$ conformality index \citep{Baltas1998}.} \\
  \mbox{Plans rated ++ are displayed in bold.}
\end{table}
\end{landscape}

\begin{landscape}
\begin{table}[T]\small{\center
\caption{\label{tbl:numeval2}Treatment plan performance indicators for patient 2.}
\begin{tabular}{llr llll llll ll ll l}
\toprule
  & & & & & & & \multicolumn{4}{c}{PTV (\%)}     & \multicolumn{2}{c}{Rectum (\unit{Gy})} & \multicolumn{2}{c}{Urethra  (\unit{Gy})} & \\
\cmidrule{8-11} \cmidrule{12-13} \cmidrule{14-15}
Model & $N$ & Sol. time & (LD) & (QD) & Alg. 1 & COIN & $D_{90\%}$ & $V_{100\%}$ & $V_{150\%}$ & $V_{200\%}$ & $D_{10\%}$ & $D_{2cc}$ & $D_{10\%}$ & $D_{0.1cc}$ & Expert \\
   & & (min) & value & value & value & & $\geq 90\%$ & $\geq 90\%$ & $\leq 55\%$ & $\leq 20\%$ & $\leq 7.2\unit{Gy}$ & $\leq 6.7\unit{Gy}$ & $\leq 10\unit{Gy}$ & $\leq 10\unit{Gy}$ & opinion \\
\midrule
IPSA$^\ast$ & 16 & $2.2$ & $0.107$ & $203$ & $0.019$ & $0.709$ & $111.3$ & $97.6$ & $37.8$ & $11.2$ & $7.6$ & $6.9$ & $9.6$ & $9.7$ & + \\
\midrule
(LD) & 15 & $1986.6$ & $0.175$ & $198$ & $0.033$ & $0.748$ & $108.3$ & $95.9$ & $36.7$ & $12.2$ & $7.3$ & $6.6$ & $9.8$ & $9.9$ & + \\
(LD) & 16 & $593.8$ & $0.156$ & $194$ & $0.030$ & $0.724$ & $110.2$ & $96.1$ & $42.9$ & $13.9$ & $7.3$ & $6.6$ & $9.8$ & $9.9$ & + \\
(LD) & 17 & $344.6$ & $0.146$ & $192$ & $0.028$ & $0.716$ & $111.0$ & $96.6$ & $46.3$ & $15.4$ & $7.3$ & $6.7$ & $9.8$ & $9.9$ & + \\
(LD) & 18 & $520.0$ & $0.144$ & $191$ & $0.028$ & $0.717$ & $111.1$ & $96.7$ & $47.1$ & $15.6$ & $7.4$ & $6.7$ & $9.8$ & $9.9$ & + \\
(LD) & 19 & $1276.3$ & $0.137$ & $193$ & $0.028$ & $0.718$ & $110.6$ & $96.6$ & $44.9$ & $15.5$ & $7.4$ & $6.7$ & $9.8$ & $9.9$ & + \\
(LD) & 20 & $364.4$ & $0.134$ & $193$ & $0.029$ & $0.723$ & $110.4$ & $96.7$ & $43.7$ & $13.8$ & $7.4$ & $6.7$ & $9.8$ & $9.9$ & + \\
\midrule
(QD) & 15 & $22.8$ & $3.242$ & $174$ & $1.127$ & $0.557$ & $80.6$ & $74.3$ & $42.4$ & $23.2$ & $6.0$ & $5.4$ & $8.3$ & $8.5$ & - - \\
(QD) & 16 & $9.0$ & $3.186$ & $173$ & $1.105$ & $0.561$ & $80.6$ & $74.9$ & $43.4$ & $23.4$ & $6.1$ & $5.5$ & $8.3$ & $8.5$ & - - \\
(QD) & 17 & $5.3$ & $3.171$ & $172$ & $1.065$ & $0.560$ & $80.4$ & $74.8$ & $43.4$ & $24.4$ & $6.1$ & $5.5$ & $8.3$ & $8.5$ & - - \\
(QD) & 18 & $2.3$ & $2.879$ & $172$ & $1.000$ & $0.569$ & $82.2$ & $75.8$ & $42.8$ & $23.4$ & $6.1$ & $5.5$ & $8.4$ & $8.5$ & - - \\
(QD) & 19 & $1.8$ & $2.762$ & $171$ & $0.964$ & $0.574$ & $82.8$ & $76.4$ & $42.4$ & $23.3$ & $6.1$ & $5.6$ & $8.3$ & $8.5$ & - - \\
(QD) & 20 & $1.5$ & $2.747$ & $171$ & $0.906$ & $0.576$ & $82.6$ & $76.6$ & $44.0$ & $23.4$ & $6.1$ & $5.6$ & $8.3$ & $8.5$ & - - \\
\midrule
Alg. 1 & 15 & $134.0$ & $0.393$ & $186$ & $0.048$ & $0.725$ & $103.7$ & $92.6$ & $45.3$ & $14.5$ & $6.8$ & $6.2$ & $9.7$ & $9.8$ & + \\
Alg. 1 & 16 & $70.6$ & $0.362$ & $185$ & $0.043$ & $0.719$ & $104.0$ & $92.9$ & $47.5$ & $14.8$ & $6.8$ & $6.2$ & $9.7$ & $9.8$ & + \\
Alg. 1 & 17 & $41.0$ & $0.359$ & $183$ & $0.043$ & $0.700$ & $104.0$ & $92.8$ & $49.0$ & $15.2$ & $6.8$ & $6.2$ & $9.7$ & $9.8$ & + \\
Alg. 1 & 18 & $27.3$ & $0.326$ & $184$ & $0.041$ & $0.709$ & $105.9$ & $93.6$ & $48.8$ & $13.4$ & $7.0$ & $6.4$ & $9.7$ & $9.8$ & + \\
Alg. 1 & 19 & $17.1$ & $0.321$ & $184$ & $0.040$ & $0.709$ & $106.0$ & $93.5$ & $50.4$ & $13.5$ & $7.0$ & $6.4$ & $9.7$ & $9.9$ & + \\
Alg. 1 & 20 & $24.9$ & $0.321$ & $183$ & $0.040$ & $0.713$ & $105.7$ & $93.5$ & $50.6$ & $14.2$ & $7.0$ & $6.4$ & $9.7$ & $9.9$ & + \\
\midrule
(LDV) & 15 & $13$ & $1.228$ & $202$ & $4.579$ & $0.753$ & $108.6$ & $95.9$ & $39.9$ & $14.1$ & $7.0$ & $6.4$ & $10.0$ & $10.1$ & + \\
(LDV) & 17 & $15$ & $0.929$ & $205$ & $2.378$ & $0.731$ & $105.5$ & $93.8$ & $31.5$ & $10.8$ & $7.2$ & $6.5$ & $9.9$ & $10.1$ & + \\
(LDV) & 19 & $2.7$ & $1.875$ & $211$ & $16.519$ & $0.736$ & $106.5$ & $94.4$ & $36.5$ & $13.8$ & $6.8$ & $6.2$ & $9.7$ & $9.8$ & + \\
(LDV) & 20 & $2.3$ & $1.228$ & $205$ & $6.033$ & $0.739$ & $107.7$ & $95.5$ & $35.4$ & $12.2$ & $7.1$ & $6.4$ & $9.8$ & $9.9$ & + \\
(LDV)$^\ast$ & 16 & $0.3$ & $1.143$ & $203$ & $6.016$ & $0.754$ & $107.5$ & $95.6$ & $39.8$ & $11.6$ & $7.0$ & $6.4$ & $9.9$ & $10.0$ & +
\\
\bottomrule
\end{tabular}
\\
} \scriptsize{$^\ast$ Using the catheter configuration chosen by an expert planner (ALH).} \\
  \mbox{Abbreviations: $N =$ allowed number of catheters, Sol.~time $=$ solution time, COIN $=$ conformality index \citep{Baltas1998}.}
\end{table}
\end{landscape}

\begin{landscape}
\begin{table}[T]\small{\center
\caption{\label{tbl:numeval3}Treatment plan performance indicators for patient 3.}
\begin{tabular}{llr llll llll ll ll l}
\toprule
  & & & & & & & \multicolumn{4}{c}{PTV (\%)}     & \multicolumn{2}{c}{Rectum (\unit{Gy})} & \multicolumn{2}{c}{Urethra  (\unit{Gy})} & \\
\cmidrule{8-11} \cmidrule{12-13} \cmidrule{14-15}
Model & $N$ & Sol. time & (LD) & (QD) & Alg. 1 & COIN & $D_{90\%}$ & $V_{100\%}$ & $V_{150\%}$ & $V_{200\%}$ & $D_{10\%}$ & $D_{2cc}$ & $D_{10\%}$ & $D_{0.1cc}$ & Expert \\
   & & (min) & value & value & value & & $\geq 90\%$ & $\geq 90\%$ & $\leq 55\%$ & $\leq 20\%$ & $\leq 7.2\unit{Gy}$ & $\leq 6.7\unit{Gy}$ & $\leq 10\unit{Gy}$ & $\leq 10\unit{Gy}$ & opinion \\
\midrule
IPSA$^\ast$ & 14 & $2.0$ & $0.347$ & $196$ & $0.391$ & $0.706$ & $108.2$ & $95.1$ & $36.1$ & $12.7$ & $7.1$ & $6.6$ & $9.6$ & $9.7$ & + \\
\midrule
(LD) & 15 & $10.1$ & $0.120$ & $203$ & $0.011$ & $0.723$ & $112.0$ & $97.2$ & $32.0$ & $10.2$ & $7.3$ & $6.8$ & $10.1$ & $10.1$ & + \\
(LD) & 16 & $7.9$ & $0.112$ & $204$ & $0.009$ & $0.737$ & $113.0$ & $97.5$ & $32.0$ & $9.9$ & $7.3$ & $6.8$ & $10.1$ & $10.2$ & + \\
(LD) & 17 & $5.9$ & $0.106$ & $203$ & $0.009$ & $0.735$ & $113.3$ & $97.5$ & $32.4$ & $9.7$ & $7.3$ & $6.8$ & $10.1$ & $10.2$ & + \\
(LD) & 18 & $3.5$ & $0.104$ & $204$ & $0.009$ & $0.743$ & $113.0$ & $97.5$ & $31.6$ & $9.2$ & $7.3$ & $6.8$ & $10.1$ & $10.2$ & + \\
{\bf (LD)} & {\bf 19} & $\mathbf{2.7}$ & $\mathbf{0.102}$ & $\mathbf{205}$ & $\mathbf{0.009}$ & $\mathbf{0.748}$ & $\mathbf{112.3}$ & $\mathbf{97.5}$ & $\mathbf{30.6}$ & $\mathbf{9.3}$ & $\mathbf{7.4}$ & $\mathbf{6.8}$ & $\mathbf{10.1}$ & $\mathbf{10.1}$ & {\bf ++} \\
{\bf (LD)} & {\bf 20} & $\mathbf{2.9}$ & $\mathbf{0.102}$ & $\mathbf{205}$ & $\mathbf{0.009}$ & $\mathbf{0.748}$ & $\mathbf{112.3}$ & $\mathbf{97.5}$ & $\mathbf{30.6}$ & $\mathbf{9.3}$ & $\mathbf{7.4}$ & $\mathbf{6.8}$ & $\mathbf{10.1}$ & $\mathbf{10.1}$ & {\bf ++} \\
\midrule
(QD) & 15 & $1.3$ & $2.355$ & $176$ & $0.713$ & $0.613$ & $87.9$ & $81.6$ & $44.8$ & $21.9$ & $6.7$ & $6.1$ & $9.2$ & $9.4$ & - - \\
(QD) & 16 & $1.2$ & $2.081$ & $175$ & $0.619$ & $0.623$ & $89.8$ & $83.0$ & $44.8$ & $21.3$ & $6.7$ & $6.1$ & $9.3$ & $9.5$ & - \\
(QD) & 17 & $0.9$ & $2.008$ & $175$ & $0.587$ & $0.629$ & $90.4$ & $83.4$ & $45.0$ & $21.0$ & $6.7$ & $6.1$ & $9.3$ & $9.5$ & - \\
(QD) & 18 & $0.9$ & $1.998$ & $175$ & $0.580$ & $0.630$ & $90.4$ & $83.4$ & $45.1$ & $20.8$ & $6.7$ & $6.1$ & $9.3$ & $9.5$ & - \\
(QD) & 19 & $0.9$ & $1.998$ & $175$ & $0.580$ & $0.630$ & $90.4$ & $83.4$ & $45.1$ & $20.8$ & $6.7$ & $6.1$ & $9.3$ & $9.5$ & - \\
(QD) & 20 & $0.8$ & $1.998$ & $175$ & $0.580$ & $0.630$ & $90.4$ & $83.4$ & $45.1$ & $20.8$ & $6.7$ & $6.1$ & $9.3$ & $9.5$ & - \\
\midrule
Alg. 1 & 15 & $22.2$ & $0.502$ & $183$ & $0.057$ & $0.689$ & $107.2$ & $94.8$ & $50.1$ & $15.7$ & $7.2$ & $6.6$ & $10.1$ & $10.3$ & + \\
Alg. 1 & 16 & $19.1$ & $0.542$ & $184$ & $0.058$ & $0.683$ & $107.7$ & $94.8$ & $52.0$ & $13.9$ & $7.3$ & $6.7$ & $10.1$ & $10.3$ & + \\
{\bf Alg. 1} & {\bf 17} & $\mathbf{16.6}$ & $\mathbf{0.504}$ & $\mathbf{184}$ & $\mathbf{0.055}$ & $\mathbf{0.684}$ & $\mathbf{108.4}$ & $\mathbf{95.3}$ & $\mathbf{52.2}$ & $\mathbf{13.2}$ & $\mathbf{7.3}$ & $\mathbf{6.8}$ & $\mathbf{10.1}$ & $\mathbf{10.3}$ & {\bf ++} \\
{\bf Alg. 1} & {\bf 18} & $\mathbf{14.9}$ & $\mathbf{0.508}$ & $\mathbf{184}$ & $\mathbf{0.056}$ & $\mathbf{0.683}$ & $\mathbf{108.4}$ & $\mathbf{95.3}$ & $\mathbf{52.3}$ & $\mathbf{14.6}$ & $\mathbf{7.3}$ & $\mathbf{6.8}$ & $\mathbf{10.1}$ & $\mathbf{10.3}$ & {\bf ++} \\
{\bf Alg. 1} & {\bf 19} & $\mathbf{14.6}$ & $\mathbf{0.529}$ & $\mathbf{184}$ & $\mathbf{0.055}$ & $\mathbf{0.686}$ & $\mathbf{108.4}$ & $\mathbf{95.3}$ & $\mathbf{53.0}$ & $\mathbf{14.2}$ & $\mathbf{7.3}$ & $\mathbf{6.7}$ & $\mathbf{10.2}$ & $\mathbf{10.3}$ & {\bf ++} \\
{\bf Alg. 1} & {\bf 20} & $\mathbf{16.1}$ & $\mathbf{0.543}$ & $\mathbf{184}$ & $\mathbf{0.053}$ & $\mathbf{0.688}$ & $\mathbf{108.5}$ & $\mathbf{95.3}$ & $\mathbf{53.5}$ & $\mathbf{13.4}$ & $\mathbf{7.3}$ & $\mathbf{6.7}$ & $\mathbf{10.2}$ & $\mathbf{10.4}$ & {\bf ++} \\
\midrule
(LDV) & 15 & $15$ & $0.814$ & $202$ & $3.153$ & $0.707$ & $108.4$ & $94.7$ & $35.9$ & $12.3$ & $6.7$ & $6.2$ & $10.0$ & $10.1$ & o \\
(LDV) & 16 & $1.4$ & $0.426$ & $204$ & $1.835$ & $0.732$ & $111.6$ & $96.9$ & $31.2$ & $9.9$ & $7.1$ & $6.6$ & $10.1$ & $10.2$ & + \\
(LDV) & 17 & $1.6$ & $0.483$ & $202$ & $1.377$ & $0.717$ & $109.7$ & $95.9$ & $37.1$ & $10.4$ & $7.0$ & $6.4$ & $10.1$ & $10.1$ & + \\
{\bf (LDV)} & {\bf 18} & $\mathbf{1.8}$ & $\mathbf{0.482}$ & $\mathbf{204}$ & $\mathbf{1.565}$ & $\mathbf{0.733}$ & $\mathbf{109.6}$ & $\mathbf{95.7}$ & $\mathbf{33.4}$ & $\mathbf{9.9}$ & $\mathbf{6.8}$ & $\mathbf{6.3}$ & $\mathbf{10.1}$ & $\mathbf{10.2}$ & {\bf ++} \\
(LDV)$^\ast$ & 14 & $8.5$ & $0.963$ & $200$ & $3.718$ & $0.710$ & $108.7$ & $95.5$ & $39.1$ & $14.2$ & $7.2$ & $6.6$ & $10.1$ & $10.1$ & + \\
\bottomrule
\end{tabular}
\\
} \scriptsize{$^\ast$ Using the catheter configuration chosen by an expert planner (ALH).} \\
  \mbox{Abbreviations: $N =$ allowed number of catheters, Sol.~time $=$ solution time, COIN $=$ conformality index \citep{Baltas1998}.} \\
  \mbox{Plans rated ++ are displayed in bold.}
\end{table}
\end{landscape}

\section{Iterative procedure for adjusting the target dose in the (QD) model} \label{ap:alg}
\begin{algorithm}[h]
	\label{alg:iterative}
	\begin{algorithmic}
		\FOR{$i \in I$}
		  \STATE $p_i := (L_i + U_i)/2$
		\ENDFOR

		\STATE CURVAL := $\infty$
		\REPEAT
			\STATE solve the (QD) model
			\STATE OLDVAL := CURVAL
			\STATE CURVAL := (QD).value
		  \FOR{$i \in I$}
		    \IF{$(\bd{D}\textbf{t})_i < L_i$}
		      \STATE $p_i := L_i$
		    \ELSIF{$(\bd{D}\textbf{t})_i > U_i$}
		      \STATE $p_i := U_i$
		    \ELSE
		      \STATE $p_i := (\bd{D}\textbf{t})_i$
		    \ENDIF
		  \ENDFOR
		\UNTIL{OLDVAL $-$ CURVAL $\leq \epsilon$}
	\end{algorithmic}
\end{algorithm}

\section{Counter example for global convergence of the iterative procedure}
\label{sec:counterexample}
We show that Algorithm \ref{alg:iterative} does not always converge to the global optimum that we would have obtained by optimizing the (QD) model simultaneously over $\textbf{t}$, $\textbf{p}$ and $b_k$.

Consider two mutually exclusive catheter positions, each offering one dwell position. There are three calculation points; the first one has $(L_1,U_1) = (8,10)$ and the other two have $(L_2,U_2) = (L_3,U_3) = (10,15)$. All dose rates are $1$ except $d_{22} = d_{23} = 2$. In the first iteration we start with prescribed dose vector $\textbf{p}=(9$ $12.5$ $12.5)$, in which case catheter position $2$ is optimal with a dwell time of $118/18 \unit{s}$. The new prescribed dose vector becomes $\textbf{p}=(8$ $13.1$ $13.1)$. In every subsequent iteration, the dose vector gets closer to $(8$ $15$ $15)$, resulting in a total penalty of $2/9$. But the optimal dose distribution is obtained by selecting catheter position $1$, which can deliver a dose of $10 \unit{Gy}$ to all three calculation points, resulting in a total penalty of $0$.
\end{document}